# Q-learning algorithm for resource allocation in WDMA-based optical wireless communication networks

Abdelrahman S. Elgamal, Osama Z. Alsulami, Ahmad Adnan Qidan, Taisir E.H. El-Gorashi and Jaafar M. H. Elmirghani
*School of Electronics and Electrical Engineering*, *University of Leeds,* Leeds, United Kingdom

*Abstract*— Visible Light Communication (VLC) has been widely investigated during the last decade due to its ability to provide high data rates with low power consumption. In general, resource management is an important issue in cellular networks that can highly effect their performance. In this paper, an optimisation problem is formulated to assign each user to an optimal access point and a wavelength at a given time. This problem can be solved using mixed integer linear programming (MILP). However, using MILP is not considered a practical solution due to its complexity and memory requirements. In addition, accurate information must be provided to perform the resource allocation. Therefore, the optimisation problem is reformulated using reinforcement learning (RL), which has recently received tremendous interest due to its ability to interact with any environment without prior knowledge. In this paper, we investigate solving the resource allocation optimisation problem in VLC systems using the basic Q-learning algorithm. Two scenarios are simulated to compare the results with the previously proposed MILP model. The results demonstrate the ability of the Q-learning algorithm to provide optimal solutions close to the MILP model without prior knowledge of the system.

*Keywords*—Visible light communication, resource allocation, MILP, and reinforcement learning.

## I. Introduction

In the past few years, the need for energy efficient high data rate communication in indoor environments has massively increased. In future, indoor users will demand data rates that might reach tens of gigabits per second per user. Unfortunately, traditional radio-based indoor wireless communication systems are incapable of meeting these requirements due to the spectrum limitations. Visible Light Communication (VLC) is a promising technology as it can provide high data rates for multiple users as introduced in [1] – [4] with low power consumption and high reliability due to its dual functionality of illumination and data communication [5], [6]. In addition, VLC systems can provide better security in the physical layer compared to radio based wireless systems [7], [8].

The development of VLC systems has resulted in a number of techniques that can play major roles in its performance enhancement in next generation cellular networks. In [9] – [17], diversity technologies such as angle diversity receivers (ADRs) were introduced to boost the signal to noise ratio (SINR). In [18] – [26], researchers proposed different adaptation techniques such as beam angle, beam power, and beam delay adaptations to improve the performance and downlink channel capacity of VLC systems as well as reducing the impact of inter-symbol interference (ISI) resulting from multipath dispersion. Furthermore, multiple access techniques were considered for VLC systems to support multiple users, maximizing the spectral efficiency. For example the investigations included multi-carrier code division multiple access (MC-CDMA) [18], [26], non-orthogonal multiple access (NOMA) [28], and wavelength division multiple access (WDMA) [29], [30]. For uplink transmission, the researchers in [31] and [32] introduced high data rate uplink channels for VLC systems using the infrared (IR) spectrum and beam steering. Finally, resource allocation (RA) techniques have been investigated by formulating optimisation problems to allocate resource in an optimal fashion to improve the communication link capacity utilising resources such as frequency, time, power, and wavelength [33], [34].

Focussing on the resource management as it is the aim of this work, the authors in [33] studied the resource allocation problem in VLC systems deploying WDMA using mixed-integer linear programming (MILP) to provide optimal users, access points, and wavelengths assignments that maximise the total signal to noise and interference ratios (SINRs). Despite the optimality of MILP, it is not considered a practical solution due to two reasons: the first is that MILP requires full knowledge of the network which is not available in many scenarios. Secondly, MILP has high complexity which increases with the density of the network, i.e. the number of users and access points. Furthermore, it requires high memory and can take a long time to provide the optimal solution.

Reinforcement learning (RL) is an important development in machine learning. RL aims to learn and build decisions for different situations within an environment in order to maximise a certain reward without any previous knowledge of the environment [35]. It was applied recently to solve various optimisation problems for different types of communication networks such as Heterogeneous Cellular Networks (HetNets) [36], Cognitive Radio Networks (CRANs) [37], Mobile Edge Computing (MEC) [38], and Software Defined Networks (SDNs). The solution to RL-based optimisation problems addresses numerous applications including but not limited to link adaptation, power control, and resource allocation. In [39] an intelligent resource allocation scheme was introduced for integrated VLC and VLC positioning (VLCP) systems using reinforcement learning to maximise the sum-rate achieved by users. The authors in [40] proposed a reinforcement learning based time-slots allocation scheme in VLC systems with dynamic time-division multiplexing (DTDMA) with the objective of maximising the spectral efficiency.

In contrast to the work proposed in [33], in this paper, we adopt the reinforcement-learning algorithm (Q-Learning) to optimise resource allocation in WDM-VLC systems. Since RL works to maximise the long-term reward, the Q-learning agent will aim to assign users, access points, and wavelengths at a given time in order to maximise the total signal to noise and interference ratio (SINR) to all users under Quality of Service (QoS) constraints.

The remainder of this paper is organised as follows: the VLC system model is discussed in Section 2. The resource allocation problem formulation using Q-learning is introduced in Section 3. After that, the simulation setup and a discussion of the results are presented in Section 4. Finally, the conclusion are provided in Section 5.

## II. SYSTEM MODEL

We consider a VLC system operating inside a room with dimensions Width ($x$), Length ($y$), and Height ($z$) as shown in Fig. 1. The VLC system consists of $N$ access points (transmitters) located on the ceiling for the purpose of illumination and data communication and $K$ users (receivers) placed in different locations on the communication floor. Each access point is equipped with 12 RYGB laser diodes (LDs) to provide four wavelengths: red, yellow, green, and blue. By using a multiplexer, these wavelengths are mixed generating an optical signal in the form of white light as proposed in [41]. The transmitters are connected together through a controller located in the optical line terminal (OLT), which is responsible for the resource allocation process. Each access point serves multiple users, each over a different wavelength by using wavelength division multiple access (WDMA). Therefore, multi-user interference can be avoided. Each user is equipped with a single wide field of view (wFOV) receiver in order to collect, filter, and separate wavelengths and extract the data associated with the wavelength.

Since the optical signal is affected by reflections and multipath propagation, the channel is modelled based on the ray-tracing algorithm proposed and used in [42], [43]. The power considered at the receiver is collected from the direct line of sight (LOS) path and reflections up to second order as reflections (as higher order have very small received power and can be neglected based on the findings in [43]). The surfaces of the ceiling, walls, and floor are divided into small equal-size areas that act as reflecting elements, which re-transmit the signal with less power in the shape of a Lambertian pattern. Temporal resolution and computation complexity are affected by the size of these reflecting elements, where a smaller element size results in high temporal resolution at the cost of high computational complexity.

The noise $\sigma$ at the user $k \in K$, which is assigned to access point $l \in L$ and wavelength $n \in N$ is due to the receiver preamplifier noise and the power received from other access points operating at the same wavelength as represented by the dashed lines in Fig. 1. Therefore, the SINR of user $k$ assigned to wavelength $n$ of access point $l$ is given by

$$SINR_{k,l,n} = \frac{P_{k,l,n}}{\sum_{l' \in L} P_{k,l',n} + \sigma_k}. \quad (1)$$

where $P_{k,l,n}$ is the power of the desired signal and $\sum_{l' \in L} P_{k,l',n}$ represents the interference received by user $k$ due to the transmission from neighbouring access point $l'$ to the other users connected to the same wavelength $n$ as illustrated in Fig. 1, using the dotted lines.

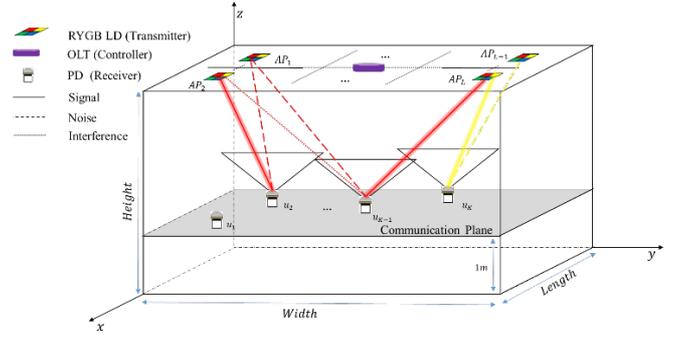

Fig. 1. System Model

## III. PROBLEM FORMULATION

The intelligent resource allocation process of VLC systems can be formulated as a Markov Decision Process (MDP) problem [44]. MDP is a mathematical scheme used to formalise decision-making problems, which are stochastic in nature and have a partly random outcome. Such problems can be solved using dynamic programming or reinforcement learning (RL) [44], [45]. In the following, our MDP model is described using four main components:

- **Agent**: Refers to the control unit responsible for the resource allocation decision.
- **State-Space $S$**: Each state $s \in S$ is a binary vector with values $\{0,1\}$ and length of $K$ users $s = \{QoS_1, \ldots, QoS_K\}$. which is defined to guarantee the minimum QoS constraint $\phi$.

$$QoS_k = \begin{cases} 1, & SINR_{k,l,n} \geq \phi_k \\ 0, & otherwise \end{cases} \forall k \in K, l \in L, n \in N. \quad (2)$$

If the minimum QoS requirements $\phi$ of a user $k \in K$ are met, $QoS_k$ equals one, otherwise, it is zero. The minimum QoS requirements will be guaranteed for all users when the state with all fields equal to one is observed.

- **Action-Space $A$**: Each action $a \in A$ describes the user, access point, wavelength assignment $x_{k,l,n}$ strategy. In this sense, $x_{k,l,n}$ is a binary value that is equal to one if user $k$ is assigned to access point $l$ and wavelength $n$. The action-space is defined to consider actions that satisfy the following constraints.

$$\sum_{n \in N} \sum_{l \in L} x_{k,l,n} = 1, \quad \forall k \in K. \quad (3)$$

$$\sum_{k \in K} x_{k,l,n} = 1, \quad \forall n \in N, \forall l \in L. \quad (4)$$

The first constraint in (3) ensures that each user is assigned to only one wavelength and one access point. While the second constraint in (4) guarantees that each wavelength within an access point is assigned to a maximum of one user.

- **Reward $R$**: As mentioned earlier, RL works by maximising a certain reward. In our VLC system, the reward will be total SINR based on the feedback from all users as in (4).

$$r(s,a) = \sum_{k \in K} SINR_{k,l,n}. \quad (5)$$

The main objective is to find the optimal policy $\pi$ that maximises the instantaneous reward $r(s,a)$. To measure the impact of following a policy, the action-value function (Q-function) is used. The Q-value $Q_\pi(s,a)$ resulting from this Q-function represents the total expected reward from taking an action $a$ when the environment is starting in state $s$. A policy $\pi^*$ is considered to be optimal if the Q-value of the policy $\pi$ converges to the optimal value.

It is worth mentioning that the agent does not know the Q-values of the policy due to its un-awareness of the environment except for the current state. Thus, all Q-values are initially set to zero. In other words, the agent cannot decide the first action in this particular state. Therefore, Q-learning was developed as an $\epsilon$-greedy algorithm to balance the exploration-exploitation trade-off where $\epsilon$ represents the exploration factor that can have a value between "0" and "1". Notice that, if the agent has no information about the environment, $\epsilon$ is set to "1" and it decreases gradually as the agent starts getting information about the environment. After that, a random value $z$ is chosen between "0" and "1". When $z > \epsilon$, the agent chooses to exploit the current Q-values, otherwise, the agent seeks to explore any unlearned possible actions. Given this point, the agent learns a new Q-value within the same policy $\pi$ for a certain state-action pair, and therefore, an update must be taken for this specific Q-value. Consequently, the updated value of the Q value is given by

$$Q_\pi^{new}(s,a) = (1-\alpha)Q_\pi(s,a) + \alpha[r(s,a) + \gamma \max_a Q_\pi(s',a)]. \quad (6)$$

where $\alpha$ is the learning rate that has a value between "0" and "1" $[0 < \alpha < 1]$. The learning rate describes the effect of the discovered new Q-value on its old value. This learning rate can be either constant along the learning process or can be tuned during the learning process. The main purpose of updating the Q-values in the Q-table is to find and minimise the temporal difference (TD) between $Q^*(s,a)$ and $Q^\pi(s,a)$ to allow convergence in the optimal Q-function. The Q-learning algorithm will abort once all Q-values within the Q-table of the policy converge to an approximate value or the number of iterations reaches a pre-set limit. This algorithm will extract the optimal policy by selecting action $a$ for a particular state $s$ that provides the maximum optimal Q-value.

$$\pi^* = \max_a Q^*(s,a). \quad (7)$$

IV. SIMULATION SETUP AND RESULTS

We consider an empty room that contains $N = 4$ lamps (access points) on the ceiling providing illumination and data communication to $K = 4$ users distributed on the communication plane (1m above the floor). Two scenarios are considered in this work; each scenario represents a different user distribution. Table 1 shows the general system configuration and the locations associated with the following scenarios:

TABLE I. SIMULATION SETUP

| Simulation Parameters | |
|---|---|
| Room Dimensions $(x, y, z)$ | $4m \times 4m \times 3m$ |
| Walls and Ceiling reflection coefficient $(\rho)$ | 0.8 |
| Floor reflection coefficient $(\rho)$ | 0.3 |
| Number of reflections | Up to $2^{nd}$ order |
| Area of reflecting element ($1^{st}$ order reflections) | $5\ cm \times 5\ cm$ |
| Area of reflecting element ($2^{nd}$ order reflections) | $20\ cm \times 20\ cm$ |
| Half-power semi-angle of reflecting elements | $60°$ |
| *Transmitter Parameters* | |
| Number of RYGB LDs per AP | 12 |
| Transmitter optical power of each wavelength in each RYGB LD. | $0.8\ W$ (Red), $0.5\ W$ (Yellow), $0.3\ W$ (Green), $0.3\ W$ (Blue) |
| Total transmitted power of each RYGB LD. | $1.9\ W$ |
| Transmitter Locations | (1,1,3), (1,3,3), (3,1,3), (3,3,3) |
| *Receiver Parameters* | |
| Photodetector FOV | $40°$ |
| Photodetector Bandwidth | $5\ GHz$ |
| Noise spectral density | $4.47\ pA/\sqrt{Hz}$ [4] |
| Photodetector area | $20\ cm^2 \times 20\ cm^2$ |
| Responsivity of each wavelength | $0.4\ A/W$ (Red), $0.35\ A/W$ (Yellow), $0.3\ A/W$ (Green), $0.2\ A/W$ (Blue) |
| Receivers Locations (scenario 1) | (1,1,1), (1,3,1), (3,1,1), (3,3,1) |
| Receiver Locations (scenario 2) | (3.5,3.5,1), (3.5, 2.5,1), (2.5,3.5,1), (2.5,2.5,1) |

- *Scenario 1:* Each user is located under a certain access point (considered as the best-case scenario).
- *Scenario 2*: All users are placed below the same access point (considered as the worst-case scenario)

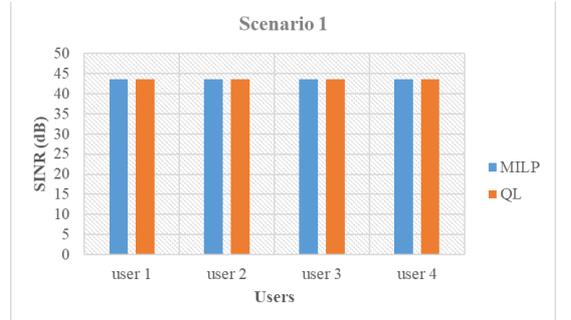

Fig. 2. SINR per user in Scenario 1.

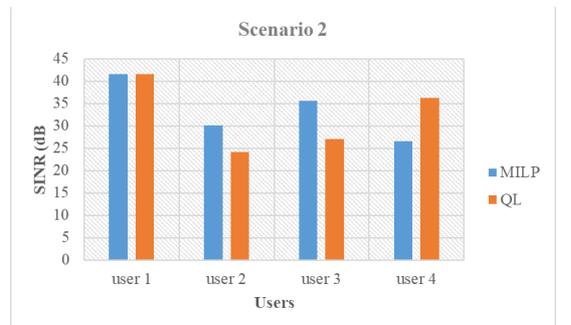

Fig. 3. SINR per user in Scenario 2.

The resource allocation optimisation problem is solved using the Q-learning algorithm and compared to the output of the MILP solution proposed in [33]. It is shown that the Q-learning algorithm can achieve sub-optimal solutions close to the optimal solutions produced from the MILP as shown in Figs. 2, 3 and 4. In Fig. 2, the best scenario is considered, and it can be seen that Q learning can converge to the optimal solution provided using the MILP model. On the other hand, the results associated with the user distribution in the worst scenario is depicted in Fig. 3. Using Q-learning, user 1 is assigned to the red wavelength, which has the highest transmission power similar to the MILP solution. While, the other users are assigned to wavelengths different than those chosen by the MILP solution. However, the sum SINR of the network is close to the optimal sum SINR obtained by the MILP as demonstrated in Fig. 4. It is worth mentioning that the output of the Q-learning algorithm is achieved without any prior knowledge of the network compared to the MILP that requires full global information.

To conclude, the results demonstrate that Q-learning can achieve a good sub-optimal solution without prior knowledge of the system while the MILP model requires full information of the system. However, the memory and time requirements are still as complex as in the MILP solution. As future work, various reinforcement-learning agents such as deep reinforcement learning and actor critic agents will be studied taking into consideration the enchantment in terms of time and memory requirements. In addition, more VLC related optimisation problems that utilise other resources such as time, frequency, and power will be formulated and solved using these advanced RL techniques.

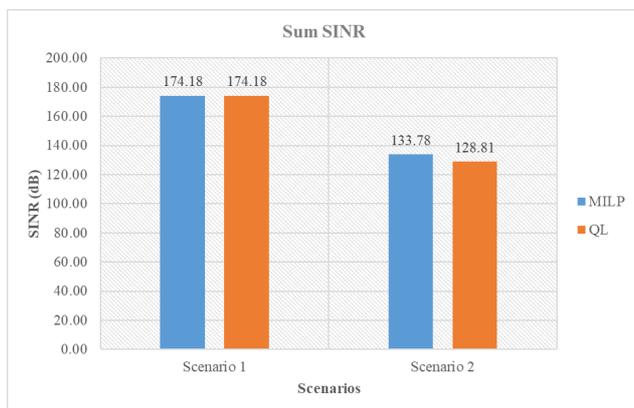

Fig. 4. Sum SINR for all user considering the two Scenarios 1 and 2.

## V. CONCLUSIONS

In this paper, the Q learning algorithm is implemented for resource allocation in a VLC network. We first formulate an optimization problem with the aim of maximizing the SINR of the network by assigning users to optimal APs and wavelengths. Then, the Q learning algorithm is used to provide a solution, which turns out to be significantly close to the optimal solution produced by the MILP in [33] avoiding complexity. The results demonstrate the ability of Q learning in providing optimal user assignment considering a uniform distribution for users on the communication floor. Moreover, Q-learning provides an acceptable solution in an environment where all users are closed to each other under a random AP. As future work, more advanced reinforcement learning algorithms will be considered in solving various optimisation problems in different contexts in VLC networks.

ACKNOWLEDGMENTS

This work has been supported in part by the Engineering and Physical Sciences Research Council (EPSRC), in part by the INTERNET project under Grant EP/H040536/1, and in part by the STAR project under Grant EP/K016873/1 and in part by the TOWS project under Grant EP/S016570/1. All data are provided in full in the results section of this paper. The first author would like to acknowledge EPSRC for funding his PhD scholarship.